\documentclass[aps,pra,floatfix,twocolumn,superscriptaddress]{revtex4}
\usepackage{amsmath,amssymb}
\usepackage{graphicx}
\usepackage{epsfig}
\usepackage{psfrag}
\usepackage[usenames]{color}
\usepackage{xcolor}
\usepackage{hyperref}
\usepackage{ulem}

\newcommand*{\colorboxed}{}
\def\colorboxed#1#{%
  \colorboxedAux{#1}%
}

\begin{document}

\title{Acoustomagnetoelectric effect in two-dimensional materials:\\ 
Geometric resonances and Weiss oscillations}


\author{I.~G.~Savenko}
\affiliation{Center for Theoretical Physics of Complex Systems, Institute for Basic Science (IBS), Daejeon 34126, Korea}
\affiliation{Rzhanov Institute of Semiconductor Physics, Siberian Branch, Russian Academy of Sciences, Novosibirsk, 630090 Russia}

\author{A.~V.~Kalameitsev}
\affiliation{Rzhanov Institute of Semiconductor Physics, Siberian Branch, Russian Academy of Sciences, Novosibirsk, 630090 Russia}

\author{L.~G.~Mourokh}
\affiliation{Physics Department, Queens College of the City University of New York, Flushing, NY
11367, USA}

\author{V.~M.~Kovalev}
\affiliation{Rzhanov Institute of Semiconductor Physics, Siberian Branch, Russian Academy of Sciences, Novosibirsk, 630090 Russia}
\affiliation{Novosibirsk State Technical University, Novosibirsk, 630072 Russia}

\date{\today}

\begin{abstract}
We study electron transport in two-dimensional materials with parabolic and linear (graphene) dispersions of the carriers in the presence of surface acoustic waves and an external magnetic field using semiclassical Boltzmann equations approach. 
We observe an oscillatory behavior of both the longitudinal and Hall electric currents as functions of the surface acoustic wave frequency at a fixed magnetic field and as functions of the inverse magnetic field at a fixed frequency of the acoustic wave. 
We explain the former by the phenomenon of geometric resonances, while we relate the latter to the Weiss-like oscillations in the presence of the dynamic superlattice created by the acoustic wave. Thus we demonstrate the dual nature of the acoustomagnetoelectric effect in two-dimensional electron gas.
\end{abstract}
%
%
%


\maketitle


\section{Introduction}
Two-dimensional (2D) electronic systems have attracted great interest of researchers for several recent decades. 
Initially, two-dimensional
electron gas (2DEG) was realized in the inversion layer at the
interface of two semiconductors with different
bandgaps~\cite{AFS}. 
Subsequently, other structures based on
graphene~\cite{graphene, 2Dgraph} and metal
dichalcogenides~\cite{2DMet} were created. 
One of the primary motivations to design a system containing 2DEG is that it represents an
ideal platform for the studies of magnetotransport which led to the
observations of quantum Hall~\cite{QHall} and fractional quantum
Hall~\cite{FHall, FHallTheor} effects.

Other prominent phenomenology is
related to magneto-oscillations of various types. Some of them are
connected to quantum effects at relatively high magnetic fields
when the Landau quantization causes the Shubnikov--de Haas effect
and associated oscillations~\cite{BeenVH}. Quantum
interference between trajectories gives rise to Aharonov-Bohm
oscillations in high-mobility GaAs/AlGaAs
heterostructures~\cite{AB}. On the other hand, semiclassical effects, which can be
observed at smaller fields or higher temperatures, are
Weiss~\cite{Weiss} and Brown-Zak (BZ) oscillations~\cite{Brown,
Zak}. The former arises due to the commensurability between the
cyclotron orbit and the spatial period in the structure, while
the latter is related to the commensurability between the magnetic
flux through the unit cell area and the magnetic flux quantum.
Subsequent Landau quantization of the BZ minibands leads to the
fractal Hofstadter Butterfly (HB) spectrum~\cite{HB}.  
Since the area of the crystal unit cell is small, it is necessary to apply extremely high fields to detect the associated phenomenology.
However, in bilayer graphene or in monolayer graphene placed on top
of a hexagonal boron nitride, additional moiré patterns appear,
which allows to observe both HB~\cite{HB1, HB2, HB3} and
BZ~\cite{BZ} oscillations.

There also exist other types of oscillations in 2D systems.
One of them is called \textit{the geometric resonances} (GRs). 
Originally, they revealed themselves in the spectra of electromagnetic power absorption coefficient of plasmas in gases and solids in the presence of a uniform magnetic field~\cite{RefGinzburg, RefCohen}.
The GRs appear as a multi-peak structure at frequencies $\omega=l\,\omega_c$, where $l$ is integer, in addition to the conventional cyclotron (or magnetoplasmon) peak at the cyclotron frequency $\omega_c=eB/m$ (or $l=1$) with $e$ and $m$ being the electron charge and mass, $B$ is the strength of the external magnetic field. 
The GRs in 2D systems have been studied theoretically~\cite{RefAndo, RefChaplik1984} and reported experimentally~\cite{RefMohr} in samples made of various materials, such as Si and AlGaAs alloys.

In this paper, we examine magnetotransport phenomena in a 2DEG in the presence of surface acoustic waves (SAWs). These waves are usually produced by the interdigital transducers (IDTs) -- metallic gates patterned on top of piezoelectric materials. 
The spacing of the gates, or pitch, determines the wavelength of the SAW~\cite{IDT}. 
When the radio-frequency (rf) signal is applied to ITDs, there emerges a SAW 
with such a wavelength that its product with the rf frequency equals to the sound velocity of the material. 
Corresponding piezoelectric field modulates both the electron density and velocity of the charge carriers. 
Accordingly, the electric current density, which is the product of these two parameters, acquires a constant component, called the acoustoelectric current. 
It can also be explained as a result of SAW drag of the charge carriers in the direction of the SAW wave vector~\cite{Parmenter}. 
The information obtained by measurements of the SAW-induced effects is complementary to conventional transport experiments, facilitating a frequent use of SAWs in the studies of low-dimensional electronic structures~\cite{SAW},  including graphene monolayers~\cite{graphene1, graphene2, OurPRLAEE}, topological insulators~\cite{top1}, and other thin films~\cite{top2}. Besides, SAWs-related methods can also be applied to the exciton transport~\cite{exciton1, exciton2, exciton3}.

The response of an electron-exposed-to-SAWs system to an external magnetic field was also examined in Refs.~\cite{RefWixforth, RefWillet, Kreft1, Kreft2}, although these studies were focused on the quantum regime with established Landau levels. 
A region of smaller fields was considered in Refs.~\cite{Levinson, Eckl} but the manifestations of Weiss oscillations were only predicted for the first-order effects, such as the SAW absorption and the velocity shifts. 
The longitudinal component of the acoustoelectric current was discussed in Ref.~\cite{Fal'ko}. 
Here, we extend this analysis to the Hall component and also examine the peculiarities appearing in the case of the linear dispersion of graphene. 

The acoustoelectric current is the second-order effect with respect to the SAW-induced electric field. Consequently, it is related to a third-order conductivity tensor~\cite{Glazov, RefBasov}. This tensor couples components of the drag current to the components of the SAW piezoelectric field as $j_\alpha=\chi_{\alpha\beta\gamma}E_\beta E_\gamma$, where $\alpha,~\beta,~\gamma=x,~y,~z$, similar to the photovoltaic effect~\cite{OurRecPRB}. As the SAW frequency is much smaller than the frequencies of the optical fields reported in Refs.~\cite{RefAndo, RefChaplik1984, RefMohr}, GRs can be expected at much smaller magnetic fields, at which a semiclassical approach based on the Boltzmann equations is appropriate for our studies. 

We calculate both the longitudinal and Hall current densities as functions of the SAW frequency and the magnetic field for two possible cases of (i) the parabolic dispersion (for the 2DEG of an interface inversion layer or of a transition metal dichalcogenide) and (ii) the linear dispersion of graphene, and we obtain an oscillatory behavior of these dependencies. We analyze these oscillations and argue that in the case of the SAW drag, GRs and Weiss oscillations represent the same phenomenon; although originally GRs are related to the optical fields with no {\it spatial} periodicity and Weiss oscillations are usually connected with a {\it static} embedded superlattice. 
SAWs thus provide a dynamical superlattice merging the GRs and Weiss oscillations phenomena and making both interpretations possible.


\section{Theoretical framework}
We start with the Boltzmann equation for the electron distribution function $f$, when the system is subject to both the piezoelectric field of the SAW and the external uniform magnetic field perpendicular to the 2D layer. 
In the case of the parabolic electron dispersion, the Boltzmann equation has the form
\begin{eqnarray}
\label{EqBolzmann}
\left[\frac{\partial}{\partial t}+\textbf{v}\frac{\partial}{\partial \textbf{r}}+e\Bigl(\textbf{E}(\textbf{r},t)+\textbf{E}^{i}(\textbf{r},t)\Bigr)\right.\\\nonumber
\left.+e[\textbf{v}\times \textbf{B}]\frac{\partial}{\partial \textbf{p}}\right]f=-\frac{f-\langle f\rangle}{\tau},
\end{eqnarray}
where 
$\mathbf{v}=\mathbf{p}/m$ is a velocity of a particle (thus the energy spectrum is given by $\varepsilon_\textbf{p}=\textbf{p}^2/2m$),
$\mathbf{r}$ is the coordinate, and $\tau$ is an effective electron scattering time. 
SAWs produce the in-plane component of a piezoelectric field $\textbf{E}(\textbf{r},t)$ directed along the SAW wave vector $\mathbf{k}$, $\textbf{E}(\textbf{r},t)||\textbf{k}$. 
$\textbf{E}^{i}(\textbf{r},t)$ is the induced field due to the spatial modulation of 2D electron density in SAW field, which can be found from the solution of the Maxwell's equation. $\langle f\rangle$ is a quasi-equilibrium electron distribution function in the SAW reference frame. This function depends on time and coordinates via the chemical potential $\mu(\textbf{r},t)$, which determines the electron density $n(\textbf{r},t)$ in slow-varying SAW field.

To find the acoustoelectric current, we expand the electron density and the distribution functions up to the second-order with respect to the total electric field $\tilde{\textbf{E}}(\textbf{r},t)=\textbf{E}(\textbf{r},t)+\textbf{E}^i(\textbf{r},t)$. In particular, $f(\textbf{r},t)=f_0+f_1(\textbf{r},t)+f_2(\textbf{r},t)+o(f_3)$, where $f_0$ is the equilibrium electron distribution function. 
The first-order correction to $f_0$ is 
$f_1(\textbf{r},t)=\left[f_1\exp(i\textbf{k}\cdot\mathbf{r}-i\omega t)+f_1^*\exp(-i\textbf{k}\cdot\mathbf{r}+i\omega t)\right]/2$, where $\omega=s|\mathbf{k}|=sk$, with $s$ being the sound velocity.

The time-independent acoustoelectric current can be determined from the stationary second-order correction to the electron distribution function $f_2$ with respect to the SAW field $\textbf{E}(\textbf{r},t)$, as
\begin{eqnarray}
\label{EqCurGeneral}
\mathbf{j}=e\int \frac{d\textbf{p}}{(2\pi\hbar)^2}\textbf{v}f_2.
\end{eqnarray}
Furthermore, we consider 2DEG to be highly degenerate, thus all the parameters are taken at the Fermi energy.
\begin{figure}[!t]
\includegraphics[width=0.49\textwidth]{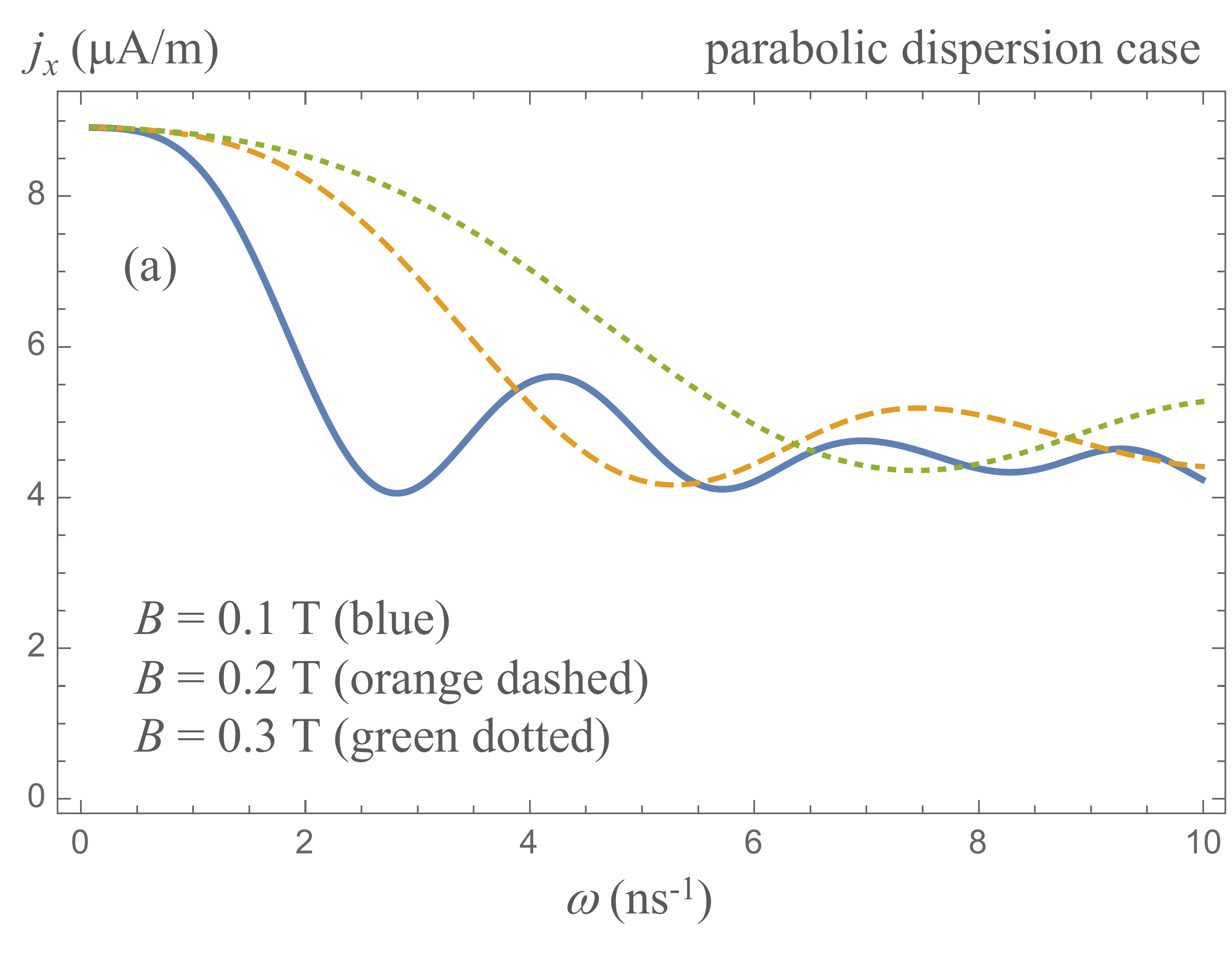}
\includegraphics[width=0.49\textwidth]{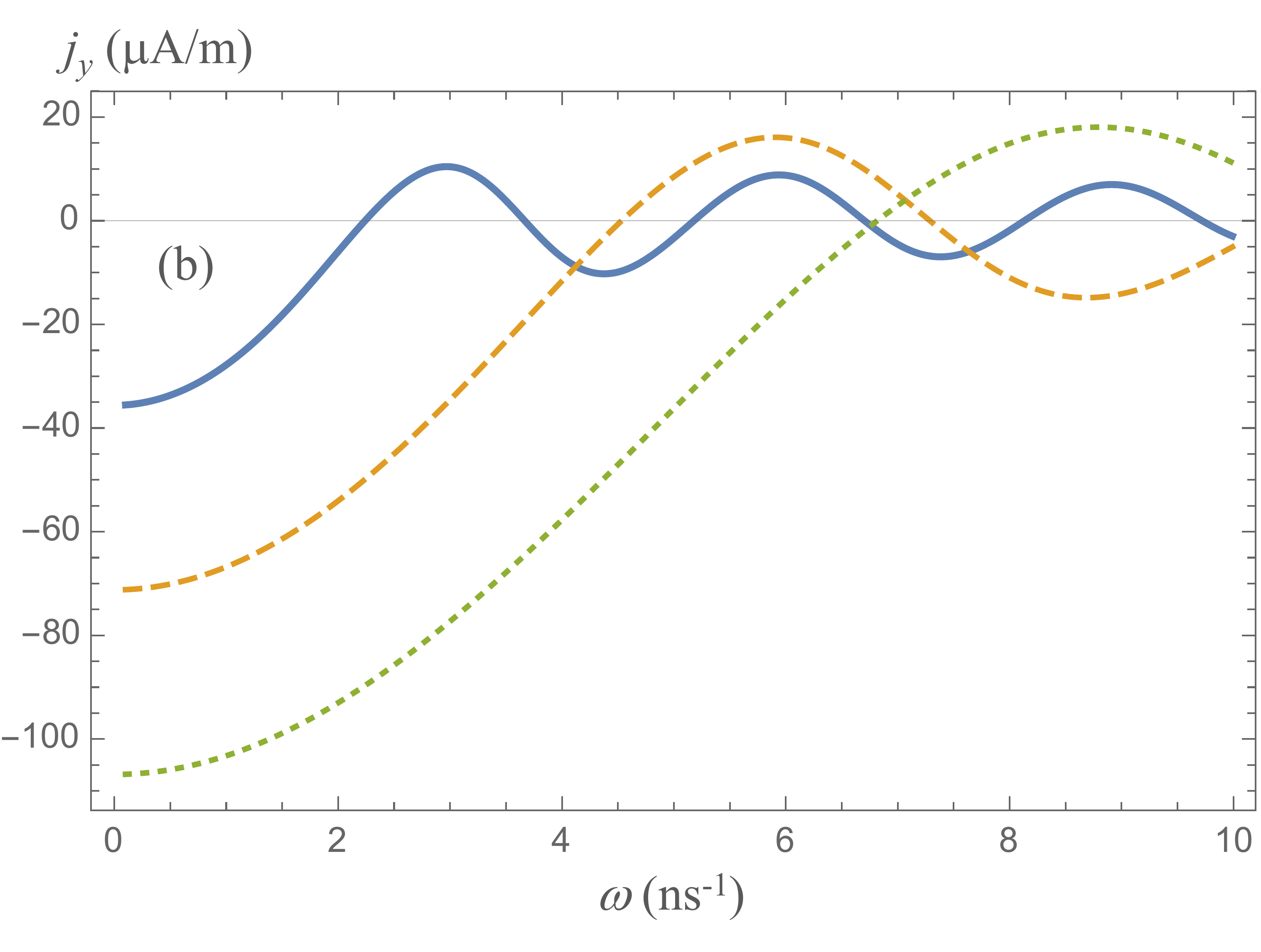}
\caption{(Color online) Electric current densities as functions of the SAW frequency for the parabolic dispersion case. (a) Longitudinal drag ($x$-component) and (b) Hall drag ($y$-component). Different colors correspond to different values of the applied magnetic field $B$, specified in panel (a).}
\label{Fig2}
\end{figure}
The $x$-axis is chosen along the direction of the SAW propagation.
After the calculations detailed in Appendix~\ref{AppendixA} and Appendix~\ref{AppendixB}, Sec.~a, we obtain the longitudinal and Hall acoustoelectric currents in the parabolic electron dispersion case, as
\begin{eqnarray}
\label{EqAux4}
&&\left(
  \begin{array}{c}
    j_x \\
    j_y \\
  \end{array}
\right)=\frac{1}{env_F}\left|\frac{\sigma_0E_0}{g(k,\omega)}\right|^2\frac{1}{\beta_F^2(1+\omega_c^2\tau^2)}\\
\nonumber
&&~~~~~~~\times \textmd{Re}\,\sum_l\frac{J_l(\beta_F)}{1-i(\omega-l\,\omega_c)\tau}\left[l+\frac{ka_0}{\omega_c\tau}\frac{\sigma_{xx}}{\varepsilon_0(s-R_x)}\right]\\
\nonumber
&&~~~~~~~\times\left(
  \begin{array}{c}
    \gamma(l+1)J_{l+1}(\beta_F)+\gamma^*(l-1)J_{l-1}(\beta_F) \\
    i\gamma(l+1)J_{l+1}(\beta_F)-i\gamma^*(l-1)J_{l-1}(\beta_F) \\
  \end{array}
\right),
\end{eqnarray}
where $\sigma_0=e^2n\tau/m$ is a static Drude conductivity, $E_0$ is the amplitude of the (external) piezoelectric field, and $J_l(\beta_F)$ are the ordinary Bessel functions with $\beta_F=kv_F/\omega_c$. 
We have also introduced two auxiliary parameters, $\gamma=1+i\omega_c\tau$ and  $a_0=2\pi\hbar^2\varepsilon_0/me^2$.
The $xx$-component of the conductivity tensor $\sigma_{xx}$ and $x$-component of the generalized diffusion coefficient $R_x$ are given by

\begin{eqnarray}
\label{EqCondxx2}
\sigma_{xx}=\frac{2\sigma_0}{\beta_F^2}\sum_l\frac{l^2J^2_l(\beta_F)}{1-i(\omega-l\,\omega_c)\tau}
\end{eqnarray}
and
\begin{eqnarray}
\label{EqCondxx3}
R_x=\frac{\omega_c}{k}\sum_l\frac{l\,J^2_l(\beta_F)}{1-i(\omega-l\,\omega_c)\tau},
\end{eqnarray}
respectively, where
\begin{eqnarray}
\label{EqDielFun}
g(k,\omega)=1+i\frac{1}{\epsilon_0(\epsilon_d+1)} \frac{\sigma_{xx}}{(s-R_x)}
\end{eqnarray}
is the dielectric function of 2DEG, $\epsilon_0$ is the dielectric permittivity of free space, and $\epsilon_d$ is the dielectric constant of the substrate. 
The function of Eq.~\eqref{EqDielFun} describes the screening of SAW piezoelectric field by the mobile electrons of 2D system.


%
%
%
\begin{figure}[!t]
\includegraphics[width=0.49\textwidth]{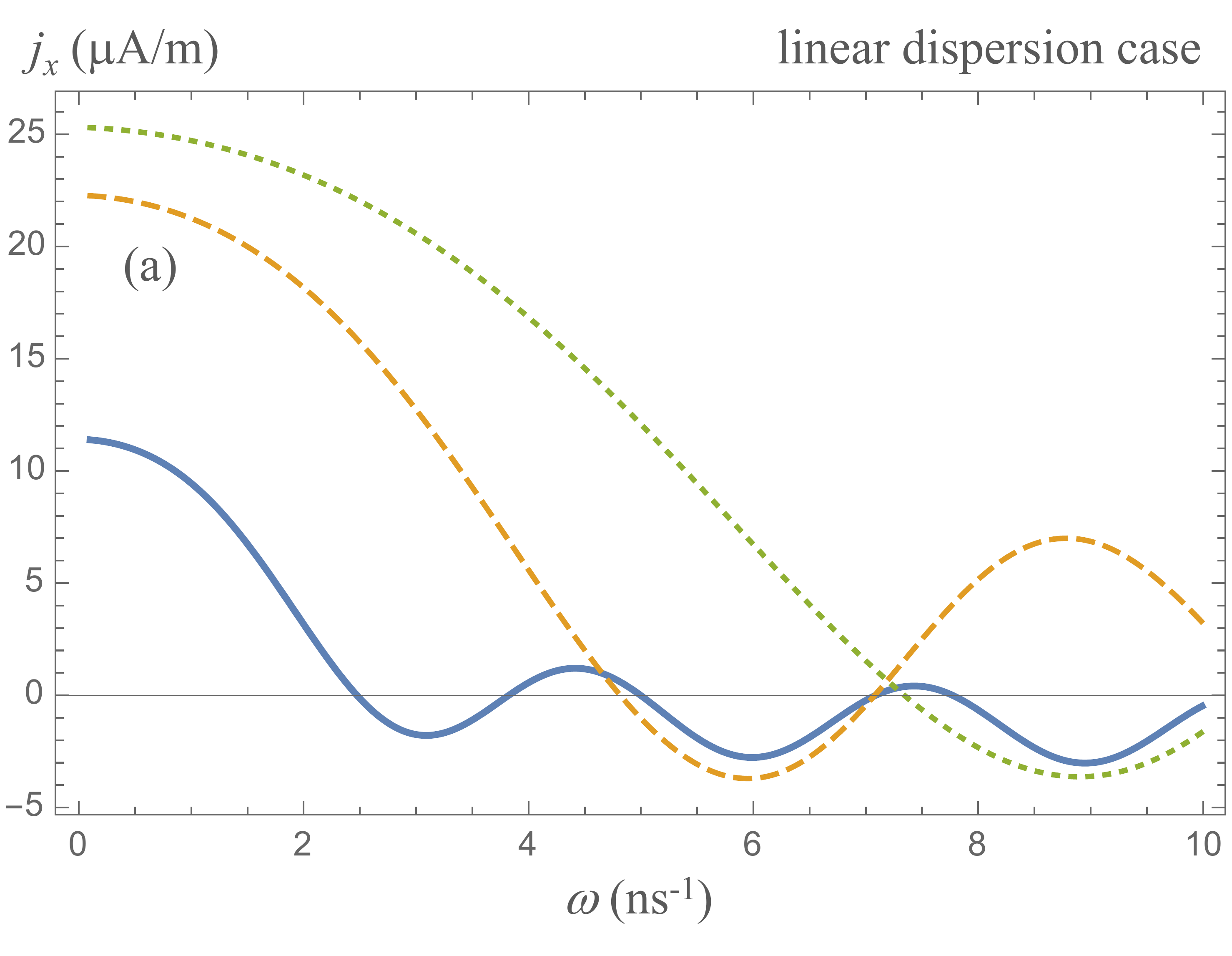}
\includegraphics[width=0.48\textwidth]{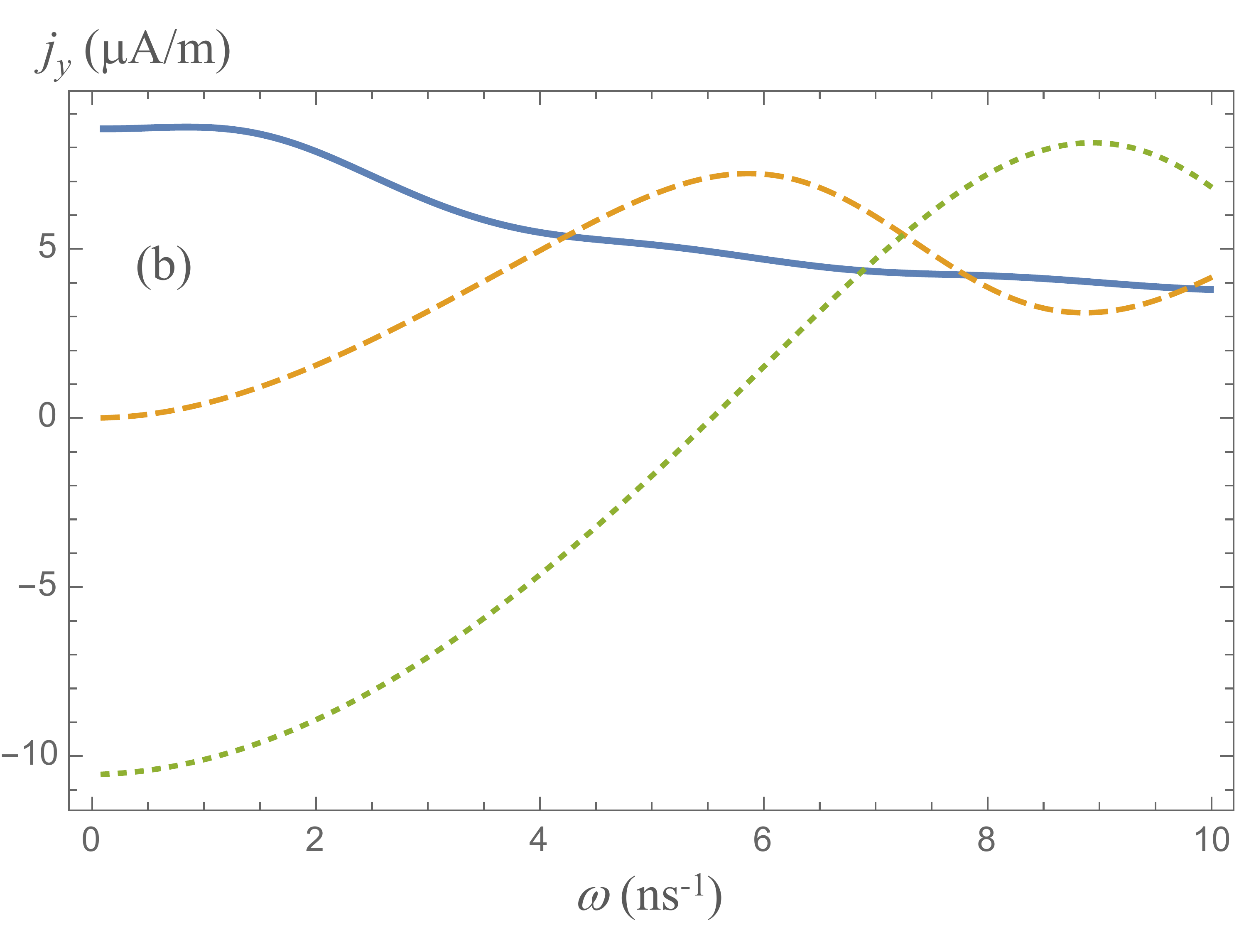}
\caption{(Color online) Electric current densities as functions of the SAW frequency for the liner dispersion case. (a) Longitudinal drag ($x$-component) and (b) Hall drag ($y$-component). Different colors correspond to different values of the applied magnetic field $B$, specified in Fig.~\ref{Fig2}(a).}
\label{Fig3}
\end{figure}
In the case of linear electron spectrum, $\varepsilon_\textbf{p}=v_0p$, 
the Boltzmann equation remains almost the same as Eq.~\eqref{EqBolzmann} with the number of changes. First, velocity $\mathbf{v}$ is replaced by $v_0\mathbf{p}/p$. 
Second, even for short-range impurities, the scattering times of the first and second harmonics of electron distribution function become energy-dependent, as $\tau_{1}(p)\equiv\tau_1(\varepsilon_\textbf{p})=\tau \varepsilon_F/\varepsilon_\textbf{p}$ for the first harmonics, and $\tau_2(p)=\tau_1(p)/2$ for the second harmonics~\cite{RefNalitov}. 
Third, the effective cyclotron frequency in the semiclassical limit is given by $\omega_c(p)=eBv_0/p=eBv_0^2/\varepsilon_p$~\cite{RefWitowski, RefOrlita}. 

Performing the calculations (see Appendix~\ref{AppendixB}, Sec.~b), we obtain the longitudinal and Hall acoustoelectric current densities in the linear electron dispersion case, as
\begin{eqnarray}
\label{MainAEGraph}
&&\left(
  \begin{array}{c}
    j_x \\
    j_y \\
  \end{array}
\right)=\frac{1}{2env_0}
\left|\frac{\sigma_gE_0}{g(k,\omega)}\right|^2
\left(\frac{1/\beta_{p_F}}{1+\omega_c^2(p_F)\tau^2_2(p_F)}\right)^2
\\
\nonumber
&&~\times
\textmd{Re}\,\sum_l\frac{J_l(\beta_{p_F})}{1-i[\omega-l\,\omega_c(p_F)]\tau_1(p_F)}\\
\nonumber
&&~\times
\left[l+\frac{ka_g}{\omega_c(p_F)\tau_1(p_F)}\frac{\sigma_{xx}}{\varepsilon_0(s-R_x)}\right]\\
\nonumber
&&~\times
\left(
  \begin{array}{c}
    -i\bar{\gamma}^2(l+1)J_{l+1}(\beta_{p_F})+i\bar{\gamma}^{*2}(l-1)J_{l-1}(\beta_{p_F}) \\
    \bar{\gamma}^2(l+1)J_{l+1}(\beta_{p_F})+\bar{\gamma}^{*2}(l-1)J_{l-1}(\beta_{p_F}) \\
  \end{array}
\right),
\end{eqnarray}
where $\sigma_g=e^2nv_0\tau_1(p_F)/p_F$ is a static Drude conductivity in graphene and all the momentum-dependent quantities are taken at $p=p_F$. In particular, $\bar{\gamma}=1+i\omega_c(p_F)\tau_2(p_F)$ and $a_g=2\pi\hbar^2\varepsilon_0v_0/e^2p_F$.
%
In this case, the $xx$-component of the conductivity tensor and $x$-component of the generalized diffusion coefficient have the forms
\begin{eqnarray}
\label{EqCondxxGraphene}
\sigma_{xx}=\frac{2\sigma_g}{\beta_{p_F}^2}\sum_l\frac{l^2J^2_l(\beta_{p_F})}{1-i[\omega-l\,\omega_c(p_F)]\tau_1(p_F)}
\end{eqnarray}
and
\begin{eqnarray}
\label{EqRxGraphene}
R_x=\frac{\omega_c(p_F)}{k}\sum_l\frac{l\,J^2_l(\beta_{p_F})}{1-i[\omega-l\,\omega_c(p_F)]\tau_1(p_F)},
\end{eqnarray}
respectively. 
We immediately see several similarities and differences between Eqs.~\eqref{EqCondxxGraphene},\eqref{EqRxGraphene} and Eqs.~\eqref{EqCondxx2},\eqref{EqCondxx3}, which we discuss below.



\section{Results and Discussion}
First of all, we want to stress that the argument $\beta_F$ of the Bessel functions in Eqs.~\eqref{EqAux4}-\eqref{EqCondxx3} and Eqs.~\eqref{MainAEGraph}-\eqref{EqRxGraphene} is of special interest. 
On one hand, it can be expressed in terms of the ratio of frequencies, as $\beta_F=\omega v_F/\omega_c s$ (in the parabolic case), resembling the GRs.
On the other hand, $\beta_F$ represents the ratio of the space scales, as $\beta_F = k r_c = 2\pi r_c/\lambda$, where $r_c$ is the cyclotron radius, which is very similar to Weiss oscillations.

To evaluate the electric current densities given by Eqs.~\eqref{EqAux4} and~\eqref{MainAEGraph}, we use the following set of parameters:
$E_0=10$~kV/m;
$n=5\cdot10^{12}$~cm$^{-2}$, which is an experimentally achievable value~\cite{RefOrlita};
$m=0.44~m_0$, where $m_0$ is a free electron mass, and we choose MoS$_2$ as a material with the parabolic spectrum; and $\tau=10^{-10}$~s, which corresponds to moderately clean samples. 
The parameters of the piezoelectric substrate are $\epsilon_d=50$ and $s=3.5\cdot 10^3$~m/s, taken for LiNbO$_3$.   For graphene, $v_0=10^8$~cm/s and $\tau_1=\mu_ep_F/ev_0$, where $\mu_e=10^4$~cm$^2$/V$\cdot$s is the electron mobility~\cite{RefMobilityGraphene1, RefMobilityGraphene2}.

Figures~\ref{Fig2} and~\ref{Fig3} show (a) longitudinal and (b) Hall components of the drag current density as functions of the SAW frequency $\omega$ for the cases of the parabolic and linear dispersions of mobile carriers, respectively, at various values of the external magnetic field. It is evident from these figures that both components exhibit oscillations, with each maximum approximately corresponding to the geometric resonance $\omega=l\,\omega_c$. As expected, for relatively small SAW frequencies and the cyclotron frequency increasing with $B$, the GRs are pronounced at magnetic fields smaller than 1~T. At higher fields, the functions are monotonic with no GRs-related oscillations. 

\begin{figure}[!t]
\includegraphics[width=0.49\textwidth]{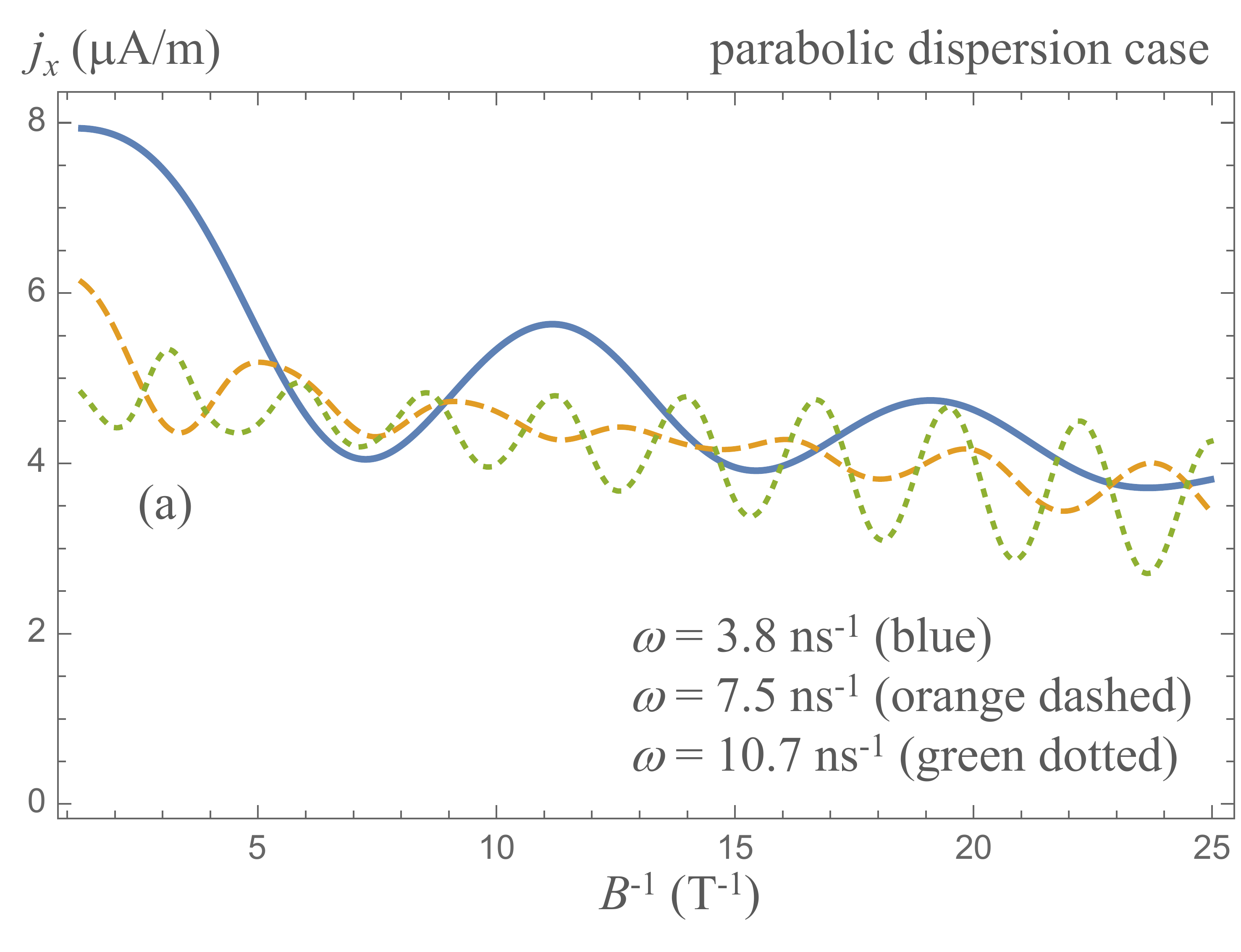}
\includegraphics[width=0.49\textwidth]{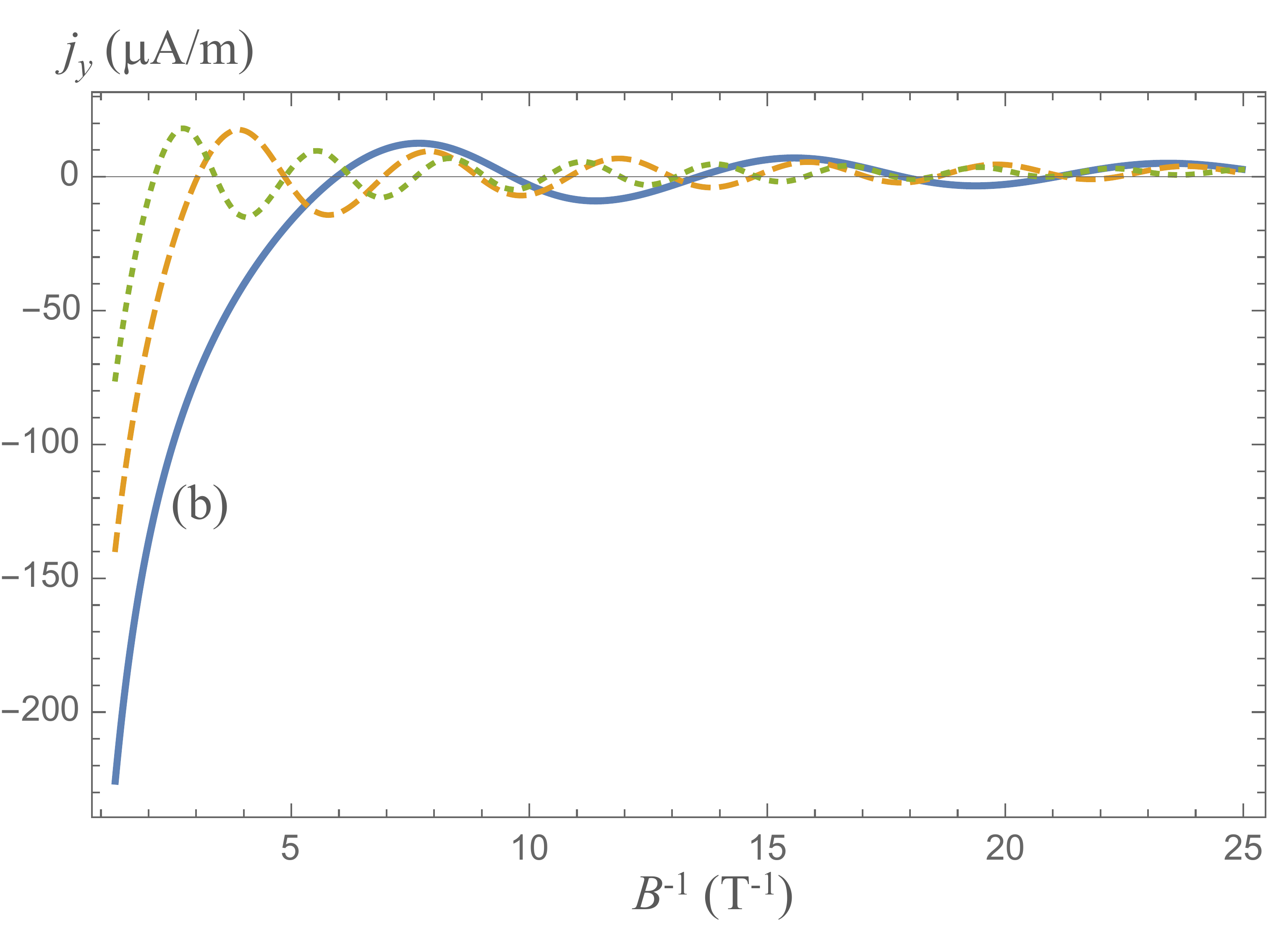}
\caption{(Color online) Components of electric current density as functions of inverse magnetic field in the case of parabolic dispersion for the frequencies specified in panel (a).
}
\label{Fig4}
\end{figure}

\begin{figure}[!t]
\includegraphics[width=0.49\textwidth]{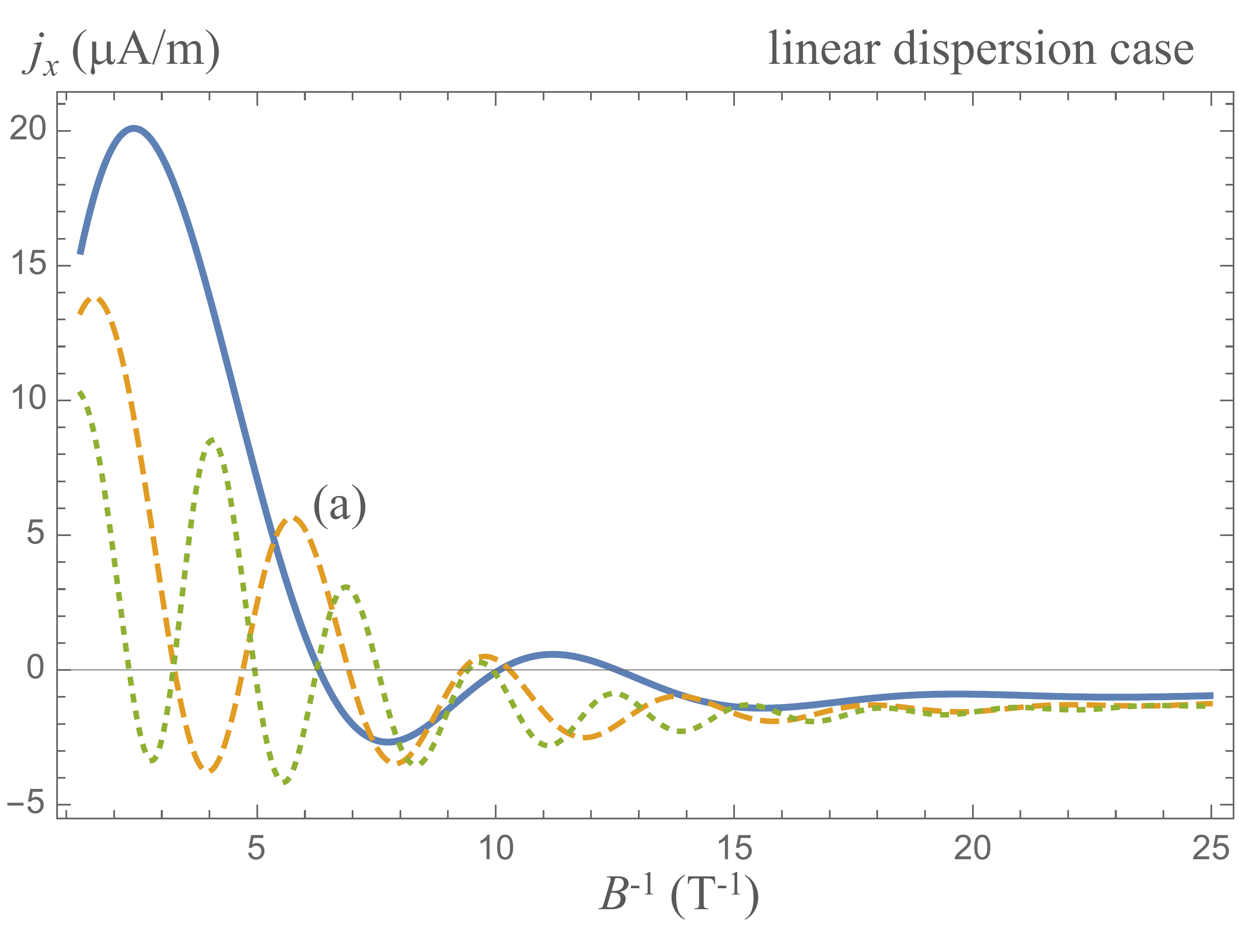}
\includegraphics[width=0.49\textwidth]{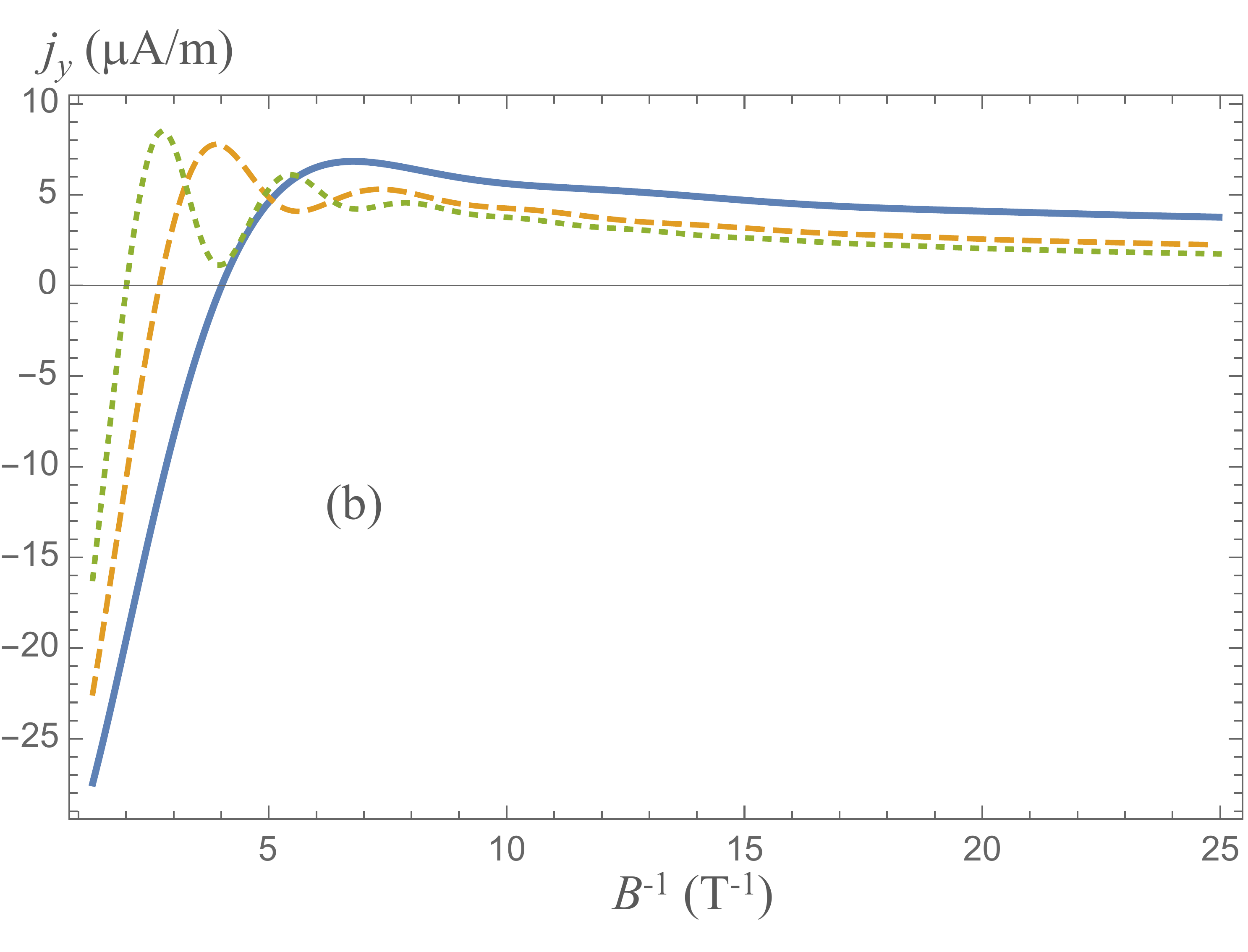}
\caption{(Color online) Components of electric current density as functions of inverse magnetic field in the case of linear dispersion for the frequencies specified in Fig.~\ref{Fig4}(a).}
\label{Fig5}
\end{figure}

The dependencies of the current density components on the inverse magnetic field are demonstrated in Figs.~\ref{Fig4} and~\ref{Fig5} for the parabolic and linear dispersion cases, respectively. 
One can see almost perfect oscillations superimposed onto the monotonic decay to the zero field. They are more pronounced for the parabolic dispersion of electrons. This result can be understood as Weiss oscillations in the presence of the spatial periodic structure of the SAW.

Another prominent feature, which we observe in the plots, is the change of the sign of the Hall current density in both the parabolic and linear dispersion cases, and the longitudinal current density in the graphene case. 
The Hall current vanishes at zero fields and monotonically increases with the increase of $B$. 
In the presence of SAW-induced oscillations of relatively high magnitude, the current density at small field can achieve negative values at minima. 
The longitudinal component of acoustoelectric current is non-zero even without a magnetic field. 
For the parabolic electron dispersion, the magnitude of the oscillations is not sufficiently large to reach negative values of the current density, while for graphene it can occur since the oscillations are more pronounced. 

It should be noted that a similar effect of the sign change was also observed in the photon drag in graphene~\cite{Ganichev1}, where it was attributed to the energy dependence of the electron scattering time. 
We believe that the same phenomenology leads to the change of the sign of the acoustoelectric current. 
We also want to emphasize that the predicted oscillating behavior of the acoustoelectric current occurs at the range of field satisfying $\hbar\omega_c\ll E_F$, where $E_F$ is the Fermi energy, validating the usage of the semiclassical approach.
%
%
%


\section{Conclusions}
To summarize, we have examined acoustoelectric current in a 2DEG in the presence of an external magnetic field in two physical systems. 
First, we have considered 2DEG in which the electron energy is proportional to its momentum squared (parabolic dispersion case). In particular, such situation occurs at the interface of two semiconductors with different band gaps and in transition metal dichalcogenides. Second, we have studied 2DEG in graphene, where the energy is proportional to the first power of momentum (linear dispersion case).

The piezoelectric field created by the SAW modulates both the electron density and electron velocity, resulting in a permanent electric current as a second-order response of the system. 
Using the semiclassical Boltzmann equations approach, we have calculated and studied both the longitudinal and Hall current densities. 
For a fixed magnetic field, both the components of the acoustoelectric current exhibit oscillations as functions of the SAW frequency. 
We have shown that the Hall component changes its sign in both cases of parabolic and linear dispersions, while the change of sign of the longitudinal component occurs in graphene only. 
For a fixed SAW frequency, the acoustoelectric current oscillates as a function of the inverse magnetic field. 

Mathematically, the oscillations are originating from the presence of the (ordinary) Bessel functions in the equations. 
The argument of Bessel functions can be represented as a ratio of the SAW and cyclotron frequencies or as a ratio of the cyclotron radius and the SAW wavelength. 
The former is conventionally used to describe optical geometric resonances, while the latter appears in Weiss oscillations of magnetoresistance in the presence of an embedded static superlattice. 
In the case of SAWs, both interpretations of this phenomenology become possible, since these two effects merge.  
\acknowledgements
We thank M. Malishava for useful discussions.
IGS acknowledges the support by the Institute for Basic Science in Korea (Project No.~IBS-R024-D1). AVK and VMK were supported by the Russian Foundation for Basic Research (Project No.~19-42-540011). LGM acknowledges the partial support by AFOSR, Award No.~FA9550-16-1-0279.



\appendix

\section{The first-order correction to the electron distribution function}
\label{AppendixA}
%
%
%
The first-order corrections to the equilibrium electron distribution function and the electron density, $f_1(\textbf{r},t)$ and
$n_1(\textbf{r},t)$, satisfy the Boltzmann equation [derived from Eq.~\eqref{EqBolzmann}],
\begin{gather}\label{EqBolzmannFirstOrder}
\left(\frac{1}{\tau}+i\textbf{k}\cdot\mathbf{v}-i\omega+e\left[\mathbf{v}\times\mathbf{B}\right]\cdot\frac{\partial }{\partial \textbf{p}}\right) f_1=\\\nonumber
=-e\Bigl(\textbf{E}+\textbf{E}^i\Bigr)\frac{\partial f_0}{\partial \textbf{p}}
\label{eq3.1main}
+\frac{n_1}{\tau}\frac{\partial f_0}{\partial n}.
\end{gather}
%
%
To find this equation, we used the expansions
\begin{gather}\label{expansion}
n(\textbf{r},t)=n+n_1(\textbf{r},t)+n_2(\textbf{r},t)+o(n_3),\\\nonumber
f(\textbf{r},t)=f_0+f_1(\textbf{r},t)+f_2(\textbf{r},t)+o(f_3),\\\nonumber
\langle f(\textbf{r},t)\rangle=f_0+[n_1(\textbf{r},t)+n_2(\textbf{r},t)+...]\frac{\partial f_0}{\partial n}+\\\nonumber
+\frac{[n_1(\textbf{r},t)+n_2(\textbf{r},t)+...]^2}{2}\frac{\partial^2 f_0}{\partial n^2}.
\end{gather}
Following the approach described in~\cite{RefLLX}, we switch to the polar system of coordinates, in which Eq.~\eqref{EqBolzmannFirstOrder} reads
\begin{gather}
\nonumber
\left(\frac{1}{\tau}-i\omega+ikv\cos\phi-\omega_c\frac{\partial }{\partial \phi}\right) f_1(p,\phi)
\\
\label{eq4.1}
=-e\tilde{E}_0v\cos\phi\frac{\partial f_0}{\partial \varepsilon_p}+\frac{n_1}{\tau}\frac{\partial f_0}{\partial n},
\end{gather}
where 
we have accounted for the fact that $\varepsilon_\mathbf{p}=\varepsilon_p$, where $p=|\mathbf{p}|$, and $\partial_\mathbf{p}f_0=(\partial_\mathbf{p}\varepsilon_p)(\partial_{\varepsilon_p}f_0)=\mathbf{v}\partial_{\varepsilon_p}f_0$ with $\partial_B A=\partial A/\partial B$.
We have also chosen the direction of $\mathbf{E}_0$ along the x-axis.
Then $\tilde{\mathbf{E}}_0=\mathbf{E}_0+\mathbf{E}^i_0$ is also directed along the x-axis (since $\mathbf{E}_0$ and $\mathbf{k}$ are collinear).
Then $\tilde{\mathbf{E}}_0\cdot\mathbf{v}=\tilde{E}_0v\cos\phi$ and $k_x=k=\omega/s$.

Eq.~(\ref{eq4.1}) can be rewritten in the form
\begin{eqnarray}\label{eq4.2}
\frac{\partial f_1}{\partial\phi}+i(\alpha-\beta\cos\phi)=Q(\phi),
\end{eqnarray}
where
\begin{eqnarray}\label{eq4.3}
\alpha=\frac{\omega+i/\tau}{\omega_c},~~~~~\beta=\frac{kv}{\omega_c}=\frac{\omega v}{s\omega_c},\\\nonumber
Q(\varphi)=\left(\frac{e\tilde{E}_0v}{\omega_c}\cos\phi+\frac{n_1}{\omega_c\tau}\frac{\partial \mu}{\partial n}\right)\frac{\partial f_0}{\partial \varepsilon_p},
\end{eqnarray}
which (all) evidently represent functions of frequency.
In Eq.~(\ref{eq4.3}), we used the relation $\partial_n f_0=(\partial_\mu f_0)(\partial_n\mu)$ and $\partial_\mu f_0=-\partial_{\varepsilon_p} f_0$, which holds for the
Fermi distribution function.

From Eqs.~(\ref{eq4.2})-(\ref{eq4.3}) we find
\begin{gather}
\label{APPEqf1Bessel}
f_1(p,\phi)=-e^{i\beta\sin\phi}\int\limits_{0}^{\infty}d\psi e^{-i\beta\sin(\phi+\psi)+i\alpha\psi}Q(\phi+\psi).
\end{gather}
Using the expansion of the exponents over the cylindrical harmonics,
\begin{gather}\label{APP2}
e^{i\beta\sin\varphi}=\sum_lJ_l(\beta)e^{il\varphi},
\end{gather}
we find
\begin{eqnarray}
\label{APP3}
&&\int\limits_0^\infty
d\psi
e^{-i\beta\sin(\phi+\psi)+i\alpha\psi}
\\
\nonumber
&&=\sum_l
 J_l(\beta)e^{il\phi}\int\limits_0^\infty e^{i(\alpha-l)\psi}d\psi
=
\sum_l
\frac{J_l(\beta)e^{-il\phi}}{i(l-\alpha)}
\end{eqnarray}
and
\begin{eqnarray}
\label{APP4}
&&\int\limits_0^\infty
d\psi
e^{-i\beta\sin(\phi+\psi)+i\alpha\psi}
\cos(\phi+\psi)
\\\nonumber
&&=
\frac{i}{\beta}\frac{\partial}{\partial\phi}
\int\limits_0^\infty
d\psi
e^{-i\beta\sin(\phi+\psi)+i\alpha\psi}
=
\frac{1}{i\beta}\sum_l
\frac{lJ_l(\beta)e^{-il\phi}}{l-\alpha}.
\end{eqnarray}
Then Eq.~\eqref{APPEqf1Bessel} transforms into
\begin{eqnarray}
\label{APPEqf1Bessel2}
f_1(p,\phi)&=&
\frac{e^{i\beta\sin\phi}}{i\omega_c}
\left(
-\frac{\partial f_0}{\partial\varepsilon_p}
\right)
\\
\nonumber
&&\times
\sum_l
\Bigl[
\frac{e\tilde{E}_0v}{\beta}l
+
\frac{n_1}{\tau}\frac{\partial\mu}{\partial n}
\Bigr]
\frac{J_l(\beta)}{l-\alpha}e^{-il\phi}.
\end{eqnarray}
%

The conductivity tensor and the diffusion vector can be calculated using the standard definition of the first-order correction to the current density,
\begin{gather}
\label{APPCurrent}
j^{(1)}_\alpha=e\int \frac{ d\mathbf{p}}{(2\pi\hbar)^2}
v_\alpha
f_1(p,\phi)=\sigma_{\alpha\beta}\tilde{E}_\beta+en_1R_\alpha,
\end{gather}
where $\mathbf{v}(\phi)=v(\cos\phi,\sin\phi)$, and
\begin{gather}\label{APPConductivity}
\sigma_{xx}=\frac{e^2}{\omega_c}\int\frac{d\mathbf{p}}{(2\pi\hbar)^2}
v^2\cos\phi~
e^{i\beta\sin\phi}\left(-\frac{\partial f_0}{\partial\varepsilon_{p}}\right)\\\nonumber
\times
\int\limits_0^\infty d\psi e^{-i\beta\sin(\phi+\psi)+i\alpha\psi}
\cos(\phi+\psi)
\end{gather}
and
\begin{gather}
\nonumber
R_x=\frac{1}{\omega_c\tau}\frac{\partial\mu}{\partial n}\int\frac{d\mathbf{p}}{(2\pi\hbar)^2}
v\cos\phi~e^{i\beta\sin\phi}\left(-\frac{\partial f_0}{\partial\varepsilon_{p}}\right)\times\\
\label{APPEqSigDiff12}
\times
\int\limits_0^\infty d\psi e^{-i\beta\sin(\phi+\psi)+i\alpha\psi}
\end{gather}
are the first ($xx$) matrix element of the conductivity tensor and the $x-$component of the diffusion vector~\cite{Chaplik, Kittel}, respectively.
Taking integrals in~\eqref{APPConductivity} and in~\eqref{APPEqSigDiff12}, we find the conductivity and the diffusion coefficient of a degenerate electron gas at zero temperature, Eqs.~\eqref{EqCondxx2} and~\eqref{EqCondxx3} in the main text.


\section{The second-order response and the AME current}
\label{AppendixB}

\subsubsection{Parabolic dispersion case}
\label{ApBssa}
Since we chose the SAW EM field to be directed along the $x$ axis, the AME current is given by the formula
\begin{eqnarray}
\label{EqCur2}
j_\alpha&=&
-\frac{e^2}{2\omega_c}\int\frac{d\mathbf{p}}{(2\pi\hbar)^2}
v_\alpha(\phi)
\int\limits_0^\infty d\psi e^{-\frac{\psi}{\omega_c\tau}}
\\
\nonumber
&&\times
\mathrm{Re}
\left\{\tilde{E}_0^*v\cos(\phi+\psi)\frac{\partial f_1(p,\phi+\psi)}{\partial\varepsilon_{p}}\right\}.
\end{eqnarray}
Expressing the $\textbf{p}$-integrals via the integrals over the energy and angle, we perform partial integrations to find
\begin{eqnarray}
\label{EqCur2}
\left(
\begin{array}{cc}
     j_x  \\
     j_y
\end{array}
\right)
&=&
\textmd{Re}\,\frac{e^2\tilde{E}_0^*}{\omega_c(2\pi\hbar)^2}\int\limits_0^\infty d\varepsilon_\textbf{p}
\int\limits_0^\infty d\psi e^{-\frac{\psi}{\omega_c\tau}}
\\\nonumber
&&\times\int\limits_0^{2\pi} d\phi
\left(
\begin{array}{cc}
     \cos\phi  \\
     \sin\phi
\end{array}
\right)\cos(\phi+\psi)
f_1(p,\phi+\psi).
\end{eqnarray}
Substituting here the first-order electron distribution function~\eqref{APPEqf1Bessel2}, we come up with the $\phi$ and $\psi$-angle integrals,
\begin{gather}\nonumber
\label{EqAux2}
\int\limits_0^{2\pi}d\phi
\cos(\phi+\psi)
\left(
\begin{array}{cc}
     \cos\phi  \\
     \sin\phi
\end{array}
\right)
e^{i\beta_F\sin(\phi+\psi)}e^{-il(\phi+\psi)}
\\
\nonumber
=
\frac{\pi}{\beta_F}
\left(
\begin{array}{cc}
     (l+1)J_{l+1}(\beta_F){e^{i\psi}}+(l-1)J_{l-1}(\beta_F){e^{-i\psi}}  \\
     i(l+1)J_{l+1}(\beta_F){e^{i\psi}}-i(l-1)J_{l-1}(\beta_F){e^{-i\psi}}
\end{array}
\right),\\\nonumber
\int\limits_0^\infty d\psi
e^{-\frac{\psi}{\omega_c\tau}{\pm i\psi}}
=
\omega_c\tau
\frac{1\pm i\omega_c\tau }{1+(\omega_c\tau)^2}.
\end{gather}
The integral over energy can be easily taken for a degenerate electrons gas, where $-\partial_{\varepsilon_p}f_0=\delta(\varepsilon_p-\mu)$.
Summing up, we find Eq.~\eqref{EqAux4} from the main text.


\subsubsection{Linear dispersion case}
\label{ApBssb}
Following similar steps as for the parabolic dispersion case,
integrating by parts via energy, and taking into account that now the cyclotron frequency and the electron relaxation time depend on energy, we find
\begin{eqnarray}
\label{EqCur2Graph}
\left(
\begin{array}{cc}
     j_x  \\
     j_y
\end{array}
\right)
&=&\textmd{Re}\,\frac{e^2\tilde{E}_0^*}{(2\pi\hbar)^2}
\\\nonumber
&\times&\int\limits_0^\infty \frac{d\varepsilon_\textbf{p}}{\omega_c(p)}
\int\limits_0^\infty d\psi e^{-\frac{\psi}{\omega_c(p)\tau_2(p)}}\left(1-\frac{\psi}{\omega_c(p)\tau_2(p)}\right)
\\\nonumber
&\times&\int\limits_0^{2\pi} d\phi
\left(
\begin{array}{cc}
     \cos\phi  \\
     \sin\phi
\end{array}
\right)\cos(\phi+\psi)
f_1(p,\phi+\psi),
\end{eqnarray}
where
\begin{eqnarray}
\label{APPEqf1BesselGraph}
f_1(p,\phi)&=&
\frac{e^{i\beta_p\sin\phi}}{i\omega_c(p)}
\left(
-\frac{\partial f_0}{\partial\varepsilon_p}
\right)
\\\nonumber
&&\times
\sum_l
\Bigl[
\frac{e\tilde{E}_0v_0}{\beta}l
+
\frac{n_1}{\tau_1(p)}\frac{\partial\mu}{\partial n}
\Bigr]
\frac{J_l(\beta_p)}{l-\alpha_p}e^{-il\phi}.
\end{eqnarray}
The integration over $\phi$ is  similar to the parabolic dispersion case, thus we find
\begin{gather}\nonumber
\label{EqAux2graph}
\int\limits_0^{2\pi}d\phi
\cos(\phi+\psi)
\left(
\begin{array}{cc}
     \cos\phi  \\
     \sin\phi
\end{array}
\right)
e^{i\beta_p\sin(\phi+\psi)}e^{-il(\phi+\psi)}
\\\nonumber
=
\frac{\pi}{\beta_p}
\left(
\begin{array}{cc}
     (l+1)J_{l+1}(\beta_p){e^{i\psi}}+(l-1)J_{l-1}(\beta_p){e^{-i\psi}}  \\
     i(l+1)J_{l+1}(\beta_p){e^{i\psi}}-i(l-1)J_{l-1}(\beta_p){e^{-i\psi}}
\end{array}
\right),
\end{gather}
whereas for $\psi$-integrals, we use
\begin{gather}
\nonumber
\int\limits_0^\infty d\psi
e^{-\frac{\psi}{\omega_c(p)\tau_2(p)}{\pm i\psi}}\left(1-\frac{\psi}{\omega_c(p)\tau_2(p)}\right)
\\\nonumber
=
\frac{\mp i\omega_c(p)\tau_2(p)}{[1\mp i\omega_c(p)\tau_2(p)]^2}.
\end{gather}
The remaining integral over energy is much simpler in the case of the degenerate electron gas due to the relation $-\partial_{\varepsilon_p}f_0=\delta(\varepsilon_p-\mu)$, using which we find Eq.~\eqref{MainAEGraph} in the main text.



\end{document}